\documentclass{jnmp}
\usepackage{graphicx}
\usepackage{amsmath}

\JNMPnumberwithin{equation}{section}

\begin{document}

\renewcommand{\evenhead}{G M Pritula and V E Vekslerchik}
\renewcommand{\oddhead}{Stationary Structures in 2D Continuous
Heisenberg Ferromagnetic Spin System}
\thispagestyle{empty}
\FirstPageHead{10}{3}{2003}{\pageref{pritula-firstpage}--\pageref{pritula-lastpage}}{Article}

\copyrightnote{2003}{G M Pritula and V E Vekslerchik}

\Name{Stationary Structures in Two-Dimensional\\
Continuous Heisenberg Ferromagnetic \\ Spin System}
  \label{pritula-firstpage}
  \Author{G~M~PRITULA~$^\dag$ and V~E VEKSLERCHIK~$^{\dag\ddag}$}
  \Address{$^\dag$~Institute for Radiophysics and Electronics,
National Academy of Sciences of Ukraine,     \\
~~Proscura Street 12, Kharkov 61085, Ukraine\\[10pt]
$^\ddag$~Departamento de Matem\'aticas, E.T.S.I. Industriales, \\
~~Universidad de Castilla-La Mancha, \\
~~Avenida de Camilo Jos\'e Cela, 3, 13071 Ciudad Real, Spain}

\Date{  Received May 30, 2002; Revised November 12, 2002;
  Accepted November 21, 2002}

\begin{abstract}
\noindent
Stationary structures in a classical isotropic two-dimensional
continuous Heisenberg ferromagnetic spin system
are studied in the framework of the $(2+1)$-dimensional
Landau--Lifshitz model. It is established that in the case of
$\vec S (\vec r, t) = \vec S (\vec r - \vec v t)$
the Landau--Lifshitz equation is closely related to the Ablowitz--Ladik
hierarchy. This relation is used to obtain soliton structures, which are
shown to be caused by joint action of nonlinearity and spatial dispersion,
contrary to the well-known one-dimensional solitons which exist due to
competition of nonlinearity and temporal dispersion. We also present
elliptical quasiperiodic stationary solutions of the stationary
$(2+1)$-dimensional Landau--Lifshitz equation.
\end{abstract}

\section{Introduction}

As it is known, despite of the fact that magnetism is an
essentially quantum effect, a wide range of magnetic phenomena can
be successfully described in the framework of classical models.
One of the most widely used of such models is the one by Landau
and Lifshitz, when magnetic is considered in the continuous limit
and interaction between magnetic dipoles is taken into account in
terms of some effective magnetic field. The simplest case of the
Landau--Lifshitz model is the case of the so-called isotropic
continuous Heisenberg ferromagnetic spin system, which is governed
by the equation
\begin{equation}
\partial_{t}\vec S = g \left[ \vec S \times \Delta\vec S \right],
\qquad
\vec S^{2} = 1.
\label{heis}
\end{equation}
Here $\partial_{t} = \partial / \partial t $ and $\Delta$ is the
two-dimensional Laplacian, $\Delta = \partial_{xx} + \partial_{yy}$.
This equation attaches much attention not only
from the viewpoint of its application in the physics of magnetic
phenomena, but also from the viewpoint of the theory of integrable
nonlinear partial differential equations. It is known that in the
$(1+1)$-dimensional case,
\begin{equation}
\partial_{t}\vec S = g \left[ \vec S \times \partial_{xx}\vec S \right],
\label{heis11}
\end{equation}
this equation, which has been discussed in a large number of
publications (see \cite{Lakshmanan,Takhtajan} as well as the books
\cite{IKK,DEGM,FT} and references
therein), can be solved using the inverse scattering transform
(IST). Another well studied reduction of (\ref{heis}) is the
static two-dimensional case, which may be referred to as
$(0+2)$-dimensional one,
\begin{equation}
\left[ \vec S \times \Delta\vec S \right] = 0,
\label{heis02}
\end{equation}
(see, e.g., \cite{DEGM}) and which is closely related to the
elliptic sine-Gordon model.

The subject of the present paper are the stationary structures of
the isotropic two-dimensional classical continuous Heisenberg spin system
and we look for solutions of (\ref{heis}) which are of the form
\begin{equation}
  \vec S = \vec S \left( x - v_{x}t, y - v_{y}t \right)
\label{stationary}
\end{equation}
(the so-called Tijon--Wright ansatz with zero frequency~\cite{Papa}).

Of course, this reduction (like any other ansatz) is a necessity,
if we want to proceed analytically, and is due to the fact that we
cannot at present integrate the original $(2+1)$-dimensional equation.

On the other hand, the stationary structures which are discussed
below are of much interest for the physics of magnetism and
nonlinear physics in general since they are realization of the
so-called dynamical solitons. In physics of magnetic phenomena
there exist two types of localized structures. First is the domain
walls (kinks) which connect two different ground states and which
cannot be destroyed without remagnetising regions of macroscopic
sizes (that is why they are called `topological solitons').
Another type is solitons, which can be viewed as bound states of a
huge number of magnons. The question of their existence and
stability, contrary to the case of the topological solitons, is
not so trivial. In some sense, they exist due to the presence of
some conserved quantities such as number of magnons, total
momentum and energy~\cite{IKK}. These are the only physical
constants of motion which do not depend on the model we use (as to
an infinite number of conservation laws appearing, e.g., in
$(1+1)$-dimensional Landau--Lifshitz equation, they seem to be an
attribute of the model and do not survive when we move to a more
realistic one). The moving stationary structures are the simplest
field configurations for which all of the constants are non-zero,
i.e. they are the simplest of general (or non-degenerate) ones.
Studying these structures one can explicitly see the interaction
of the nonlinearity and dispersion which is known to be the core
mechanism of creation of localized objects (solitons) not only in
magnetic systems but in many other areas of the nonlinear physics.

After substitution (\ref{stationary}) equation (\ref{heis}) becomes
\begin{equation}
g \left[ \vec S \times \Delta\vec S \right] +
\left( \vec v , \nabla \vec S \right) = 0.
\label{mainvec_eq}
\end{equation}
Introducing the variables
\begin{equation}
z = \frac{v}{4g}
\left[ x - v_{x}t + i( y - v_{y}t) \right],
\qquad
\bar z = \frac{v}{4g}
\left[ x - v_{x}t - i( y - v_{y}t) \right]
\end{equation}
where $v = \left| \vec v \right|$, we can rewrite it as
\begin{equation}
\left[ \vec S \times \vec S_{z \bar z} \right] +
\lambda^{2} \vec S_{z} + \lambda^{-2} \vec S_{\bar z} = 0
\label{Tijon}
\end{equation}
or, equivalently, as
\begin{equation}
\vec S_{z \bar z} +
\left( \vec S_{z} \vec S_{\bar z} \right) \vec S +
\left[
  \left(
  \lambda^{2} \vec S_{z} + \lambda^{-2} \vec S_{\bar z}
  \right)
 \times  \vec S
\right] = 0
\label{veczz}
\end{equation}
with $\lambda=\exp(i\gamma/2)$, where the angle $\gamma$ is
defined by $v_{x}=v\cos\gamma$, $v_{y}=v\sin\gamma$.
Equation~(\ref{Tijon}) is known to be integrable (its zero-curvature
representation (ZCR) one can find in the paper~\cite{Papa}), and
one can tackle it by elaborating the corresponding inverse
scattering transform. However, in the present paper we do not
discuss this question. Our aim is to establish the relations
between the model considered and the other integrable models,
which will provide us with a wide range of physically interesting
solutions.

For our further purposes it is convenient to rewrite~(\ref{veczz}) in the
matrix form using the correspondence
\begin{equation}
\vec S = \left( S_{1}, S_{2}, S_{3} \right)
\rightarrow
S = \sum_{a=1}^{3} S_{a} \sigma^{a},
\label{vec-matr}
\end{equation}
where $\sigma^{a}$ are the Pauli matrices
\begin{equation}
\sigma^{1} = \left(\begin{array}{cc} 0 & 1 \\ 1 & 0\end{array}\right), \qquad
\sigma^{2} = \left(\begin{array}{cc} 0 & -i \\ i & 0\end{array}\right), \qquad
\sigma^{3} = \left(\begin{array}{cc} 1 & 0 \\ 0 & -1\end{array}\right).
\label{pauli}
\end{equation}
Equation (\ref{veczz}) then can be presented as
\begin{equation}
S_{z \bar z} +
\frac 12
\left( {\rm tr}\, S_{z} S_{\bar z} \right) S +
\frac{ 1}{2i }
\left[
\lambda^{2} S_{z} + \lambda^{-2} S_{\bar z} \, ,  S \right]
= 0.
\label{main_eq}
\end{equation}
Namely this equation is the central object of our investigation.

We will use the following remarkable fact: equation
(\ref{main_eq}) is gauge equivalent to the $O(3,1)$ nonlinear
$\sigma$-model \cite{sigma} in the similar way as, e.g., the
$(1+1)$-dimensional classical continuous Heisenberg ferromagnetic
spin system (\ref{heis11}) is equivalent to the
nonlinear Schr\"odinger equation (see \cite{DEGM,FT}), or the
Ishimori magnetic \cite{Ishim} -- to the Davey--Stewartson system
(see~\cite{LS}). This equivalence can be briefly described in
terms of the IST as follows: some combinations of the Jost
functions of the linear problem associated with the $O(3,1)$
$\sigma$-model solve equation (\ref{main_eq}) (below we shall
discuss this question more comprehensively). The fact that model
(\ref{veczz}), or (\ref{main_eq}), is related to the $O(3,1)$
nonlinear $\sigma$-model is a genera\-li\-zation of the already known
result (see, e.g., \cite{DEGM}) that in the static case equation
(\ref{heis02}) is gauge equivalent to the elliptic sine-Gordon
equation. The $O(3,1)$ $\sigma$-model, as it has been shown
in~\cite{sigma}, is, in its turn, closely related to the
Ablowitz--Ladik hierarchy (ALH)~\cite{AL1}. So we shall establish
the direct links between the model considered and the ALH, which
is much more well-studied than equations~(\ref{mainvec_eq}),
(\ref{main_eq}) or models~\cite{Papa,sigma}.

In the present paper we first derive the gauge equivalence between
the Heisenberg equation and the ALH (Sections~\ref{sec_HeisALH},
\ref{sec_gauge}) and then use it to obtain soliton solutions
(Section~\ref{sec_sol}) and the elliptical quasiperiodic ones
(Section~\ref{sec_QPS}).

\section{The Heisenberg equation and the Ablowitz--Ladik\\ hierarchy}
\label{sec_HeisALH}

The method used here, which may be called the `embedding into the
ALH' method, has been discussed in \cite{sigma,2dtl,DS}. Its main
idea is that some equations can be, in some sense, `derived' from
the system of differential-difference equations (DDE) belonging to
the ALH, which means that any common solution of several equations
from the ALH also solves the equation we are dealing with.
Relatively to the problem considered this can be briefly outlined
as follows.

Consider the system of two equations from the ALH,
\begin{gather}
i \partial_{x} q_{n} =
p_{n} \left( q_{n+1} + q_{n-1} \right),
\label{dnlse}
\\
\partial_{y} q_{n} =
p_{n} \left( q_{n+1} - q_{n-1} \right),
\label{dmkdv}
\end{gather}
where
\begin{equation}
p_{n} = 1 - q_{n}r_{n}, \qquad
r_{n} = -\kappa \bar q_{n}, \qquad
\kappa = \pm 1.
\end{equation}
Equation (\ref{dnlse}) is the well-known discrete nonlinear
Schr\"odinger equation (DNLSE) \cite{AL2}, modified by the substitution
$q_{n} \rightarrow q_{n}\exp(2ix)$, while the next one, (\ref{dmkdv}),
is the discrete modified KdV equation (DMKdV) \cite{AS}. These equations
can be rewritten in terms of the complex variables $z = x + iy$,
$\bar z = x - iy$ as
\begin{gather}
i \partial q_{n} = p_{n} q_{n+1},
\label{1}\\
i \bar\partial q_{n} = p_{n} q_{n-1},
\label{-1}
\end{gather}
where $\partial$ stands for $\partial/\partial z$ and $\bar\partial$ for
$\partial/\partial \bar z$.

It is very important that these equations are compatible, since
they belong to the same hierarchy, and the constants of motion
that play the role of the Hamiltonians for the flows
(\ref{dnlse}), (\ref{dmkdv}) are in involution. Hence, we can
consider them simultaneously, as one system of two equations. It
has been shown in \cite{sigma}, and one can easily verify this
fact by simple calculations, that, for any fixed $n$, each
solution of system (\ref{dnlse}), (\ref{dmkdv}) also solves the
field equations of the $O(3,1)$ nonlinear $\sigma$-model,
\begin{equation}
  \partial\bar\partial \, q +
  \frac{ (\partial q)(\bar\partial q) r}{1-qr}
+ (1-qr) q = 0 ,
\qquad
r = -\kappa \bar q
\label{o31}
\end{equation}
and that the quantities $p_{n}$ satisfy the 2D Toda lattice
equations
\begin{equation}
\partial\bar\partial \,
\ln p_{n} = p_{n-1} - 2 p_{n} + p_{n+1}
\end{equation}
(see \cite{2dtl}). Namely this we bear in mind when say that the
O(3,1) nonlinear $\sigma$-model and the 2D Toda lattice can be
`embedded' into the ALH.

The situation with the Heisenberg spin system is somewhat more
difficult. Solution for equation (\ref{main_eq}) cannot be
constructed by means of $q_{n}$'s and $r_{n}$'s only. To do that
we have to consider the ZCR for the ALH and to analyze the
corresponding linear problems.

The integrable DDEs (\ref{1}), (\ref{-1}), as well as all equations of the
ALH, can be presented as the compatibility condition for the linear system
\begin{gather}
\Psi_{n+1}(\lambda) = U_{n}(\lambda) \Psi_{n}(\lambda),
\label{zcr-sp}
\\
\partial \Psi_{n}(\lambda) = V_{n}(\lambda) \Psi_{n}(\lambda),
\label{zcr-evolp}
\\
\bar\partial \Psi_{n}(\lambda) = W_{n}(\lambda) \Psi_{n}(\lambda),
\label{zcr-evoln}
\end{gather}
where
\begin{equation}
U_{n} =
\left(\begin{array}{cc} \lambda & r_{n} \\ q_{n} & \lambda^{-1} \end{array}\right)
\end{equation}
and the matrices $V_{n}$, $W_{n}$ are given by
\begin{equation}
V_{n} =
- i \left(\begin{array}{cc}
0 & \lambda^{-1} r_{n-1} \vspace{1mm}\\ \lambda^{-1} q_{n} & \lambda^{-2} -
r_{n-1}q_{n} \end{array}\right) , \qquad W_{n} = i \left(\begin{array}{cc} \lambda^{2} -
q_{n-1}r_{n} & \lambda r_{n} \vspace{1mm}\\ \lambda q_{n-1} & 0 \end{array}\right). \label{VW}
\end{equation}
One can easily see that equations (\ref{zcr-sp}),
(\ref{zcr-evolp}) and (\ref{zcr-sp}), (\ref{zcr-evoln}) are
compatible only if matrices $U$, $V$ and $W$ satisfy the so-called
zero-curvature equations
\begin{equation}
\partial U_{n} = V_{n+1} U_{n} - U_{n} V_{n}
\label{ZCR-p}
\end{equation}
and
\begin{equation}
\bar\partial U_{n} = W_{n+1} U_{n} - U_{n} W_{n}
\label{ZCR-n}
\end{equation}
which are equivalent to (\ref{1}) and (\ref{-1}) correspondingly.

Namely the solutions of the linear problems (\ref{zcr-sp})--(\ref{zcr-evoln}),
 $\Psi_{n}$'s, are the key objects of our
consideration and the main result of this paper can be formulated
as follows: for any~$n$, matrices
\begin{equation}
S_{n} =
\Psi_{n}^{-1}
\sigma^{3}
\Psi_{n}
\end{equation}
constructed of solutions of the linear problems of the ALH solve
the matrix Landau--Lifshitz equation (\ref{main_eq}).

To derive this result consider matrices $\sigma_{n}^{a}$, $a=1,2,3$
defined by
\begin{equation}
\sigma_{n}^{a} =
\sigma_{n}^{a} \left( \lambda; z, \bar z \right) =
  \Psi_{n}^{-1}( \lambda; z, \bar z)
  \sigma^{a}
  \Psi_{n}( \lambda; z, \bar z ),
\label{sigma}
\end{equation}
where $\sigma^{a}$ is a Pauli matrix (\ref{pauli}), and
$\Psi_{n}$, recall, is a matrix solution of system (\ref{zcr-sp}),
(\ref{zcr-evoln}) (in this notation $S_{n}=\sigma_{n}^{3}$). It
follows from (\ref{sigma}) and (\ref{zcr-evolp}),
(\ref{zcr-evoln}) that
\begin{equation}
\partial \sigma^{a}_{n} =
\Psi_{n}^{-1} \left[ \sigma^{a}, V_{n} \right]  \Psi_{n}
,\qquad
\bar\partial \sigma^{a}_{n} =
\Psi_{n}^{-1} \left[ \sigma^{a}, W_{n} \right]  \Psi_{n}.
\end{equation}
Using expressions (\ref{VW}) for $V_{n}$ and $W_{n}$ one can find
the derivatives of the matrices $S_{n}$ in terms of the matrices
$S^{\pm}_{n}$ given by $S^{\pm}_{n}=(1/2)\left(\sigma^{1}_{n} \pm
i\sigma^{2}_{n}\right)$ as follows:
\begin{gather}
(i\lambda / 2)\; \partial S_{n} =
  r_{n-1} S^{+}_{n} - q_{n} S^{-}_{n},
\label{dsdz}
\\
(i / 2\lambda)\; \bar\partial S_{n} =
   - r_{n} S^{+}_{n} + q_{n-1} S^{-}_{n}.
\label{dsdbz}
\end{gather}
These relations together with analogous expressions for the
derivatives $\partial\sigma^{\pm}_{n}$,
$\bar\partial\sigma^{\pm}_{n}$ and formulae (\ref{1}), (\ref{-1}),
after straightforward calculations, omitted here, lead us to
\begin{equation}
\frac{1}{2 } \partial \bar\partial S_{n} =
\left( q_{n-1}r_{n-1} + q_{n}r_{n} \right) S_{n} +
\left( \lambda^{-1} r_{n} - \lambda r_{n-1} \right) S^{+}_{n} +
\left( \lambda^{-1} q_{n-1} - \lambda q_{n} \right) S^{-}_{n}.
\end{equation}
Noting that
\begin{equation}
{\rm tr}\,
  \partial S_{n} \, \bar\partial S_{n} =
- 4 \left( q_{n-1}r_{n-1} + q_{n}r_{n} \right)
\end{equation}
and
\begin{equation}
\left[
  \lambda^{2} \partial S_{n} +
  \lambda^{-2} \bar\partial S_{n}\,,
  S_{n}
\right]
=
4i \left( \lambda r_{n-1} - \lambda^{-1} r_{n} \right) S_{n}^{+} +
4i \left( \lambda q_{n} - \lambda^{-1} q_{n-1} \right) S_{n}^{-}
\end{equation}
(both of these formulae follow from (\ref{dsdz}), (\ref{dsdbz}))
we obtain that for every $n$ the matrix~$S_{n}$ solves the
equation
\begin{equation}
\partial \bar\partial S_{n} +
\frac{1}{2 }
\left( {\rm tr}\, \partial S_{n} \bar\partial S_{n} \right) S_{n} +
\frac{1}{2i}
\left[
  \lambda^{2} \partial S_{n} +
  \lambda^{-2} \bar\partial S_{n}\, ,
  S_{n}
\right]
= 0
\end{equation}
which is the main equation of our study, (see (\ref{main_eq})).
This key result of the present paper can be reformulated in terms
of the vector $\vec S_{n}$, which corresponds to the matrix
$S_{n}$ and which can be presented as
\begin{equation}
\vec S_{n} = \left( S_{n1},  S_{n2},  S_{n3} \right),
\label{vecSn}
\end{equation}
where
\begin{equation}
S_{n1} + iS_{n2} =
  \frac{ \Psi_{n}^{(11)}\Psi_{n}^{(21)}}
{    \Psi_{n}^{(11)}\Psi_{n}^{(22)} - \Psi_{n}^{(12)}\Psi_{n}^{(21)} }
, \qquad
S_{n3} =
  \frac{ \Psi_{n}^{(11)}\Psi_{n}^{(22)} + \Psi_{n}^{(12)}\Psi_{n}^{(21)}}
{    \Psi_{n}^{(11)}\Psi_{n}^{(22)} - \Psi_{n}^{(12)}\Psi_{n}^{(21)} }
\label{vecSn123}
\end{equation}
(here $\Psi_{n}^{(ij)}$ are the elements of the matrix $\Psi_{n}$), as
follows: for each $n$ the vector $\vec S_{n}$ defined by (\ref{vecSn}),
(\ref{vecSn123}) solves equation (\ref{veczz}).

Thus we have established the links between equation
(\ref{main_eq}), or (\ref{veczz}), describing stationary moving
structures in the $(2+1)$-dimensional classical continuous
Heisenberg spin system and the ALH. Some more detailed analysis of
the gauge equivalence between these models one can find in the
next section. However, in this work we are going to focus our
attention on `practical' aspects of this relation, so a reader can
consider it as an `empirical' fact which can be straightforwardly,
and rather easily, verified by the calculations outlined above.

As was mentioned earlier, model (\ref{Tijon}) is known to be
integrable and its zero-curvature representation has already been
written out. But, to our knowledge, the corresponding IST has not
been elaborated yet, while the ALH is one of the best-studied
nonlinear integrable models. Besides, the Heisenberg equation is a
vector problem, which somehow complicates inverse scattering
analysis, while the ALH is a scalar one. So, to our opinion, the
`embedding into the ALH' approach is rather promising and in what
follows we demonstrate its usefulness by constructing the soliton
and quasiperiodic solutions for the equations considered using the
already known solutions for the ALH.

The magnetic energy density, ${\cal W}$,
\begin{equation}
{\cal W} \left[ \vec S \right] =
\frac{ g}{2} \left( \nabla \vec S, \nabla \vec S \right)
\label{W}
\end{equation}
of the field configurations obtained by the embedding into the ALH method
can be expressed in terms of the $q_{n}$ and $r_{n}$'s:
\begin{equation}
{\cal W} \left[ \vec S_{n} \right] =
- \frac{ v^{2}}{4 g}
\left( q_{n}r_{n} + q_{n-1}r_{n-1} \right).
\label{Wqr}
\end{equation}

It can be shown that from the viewpoint of application of solutions of
the ALH equations to the description of the vector field $\vec S$ one has
restrict himself with the case of $\kappa = 1$,
\begin{equation}
r_{n} = - \bar q_{n},
\qquad
p_{n} = 1 + \left| q_{n} \right|^{2}
\label{kappa}
\end{equation}
when the components of the vector $\vec S_{n}$ (\ref{vecSn}) are
real (in the opposite case, $\kappa=-1$, the components of $S_{n}$
are complex) and the magnetic energy (\ref{Wqr}) is positive.

In the next section we will consider the relation between equation
(\ref{main_eq}) and the ALH in the framework of the IST.

\section{Gauge equivalence and zero curvature representation}
\label{sec_gauge}

In the previous section we considered the relation between the ALH
and the Landau--Lifshitz equation in terms of {\it solutions}: we
demonstrated how to use solutions of the ALH to obtain ones for
the Landau--Lifshitz equation. Now we are going to discuss this
question in somewhat more general way. Both the ALH and the
Landau--Lifshitz equations are integrable models and it is
interesting to describe this correspondence in the language of the
IST and to derive links between the auxiliary linear problems
which are used to present the integrable models in the zero
curvature form (namely this is usually understood when one uses
the words `gauge equivalence').

Let us consider again the auxiliary linear problems of the ALH mentioned
in Section~\ref{sec_HeisALH}. To our current purposes we do not need the
discrete problem~(\ref{zcr-sp}) and will be dealing with the continuous
ones (\ref{zcr-evolp}), (\ref{zcr-evoln}). So, we omit now the index $n$
and rewrite (\ref{zcr-evolp}), (\ref{zcr-evoln}), (\ref{VW}) as
\begin{equation}
\partial \Psi(\zeta) = V(\zeta) \Psi(\zeta)
,\qquad
\bar\partial \Psi(\zeta) = W(\zeta) \Psi(\zeta),
\label{zcr-evol-g}
\end{equation}
where
\begin{equation}
  V(\zeta) =
  - i \left(\begin{array}{cc}
    0 & \zeta^{-1} r_{0} \vspace{1mm}\\
    \zeta^{-1} q_{1} & \zeta^{-2} - r_{0}q_{1} \end{array}\right)
\label{V-alh}
\end{equation}
and
\begin{equation}
  W(\zeta) =
  i \left(\begin{array}{cc}
    \zeta^{2} - q_{0}r_{1} & \zeta r_{1} \vspace{1mm}\\
    \zeta q_{0} & 0 \end{array}\right)
\label{W-alh}
\end{equation}
(we have replaced $q_{n}$, $r_{n}$ with $q_{1}$, $r_{1}$ and
$q_{n-1}$, $r_{n-1}$ with $q_{0}$, $r_{0}$). In what follows we
denote the spectral parameter by $\zeta$ and use $\lambda$ for its
particular value appearing in the definition~(\ref{sigma}) of the
matrix $S$,
\begin{equation}
S =
S \left(z, \bar z \right) =
  \Psi^{-1}( \lambda; z, \bar z)
  \sigma^{3}
  \Psi( \lambda; z, \bar z )
\label{sigma-g}
\end{equation}
The compatibility (zero-curvature) condition for the system
(\ref{zcr-evol-g})
\begin{equation}
\bar\partial V(\zeta) -
\partial W(\zeta) +
\left[ V(\zeta), W(\zeta) \right] = 0
\label{ZCR-g}
\end{equation}
leads to the following system of four {\it partial differential equations}
(PDE) for four unknown functions $q_{1}$, $r_{1}$, $q_{0}$, $r_{0}$:
\begin{gather}
i \partial q_{0} = p_{0} q_{1},
\label{q0-g}\\
i \partial r_{1} = - p_{1} r_{0},
\label{r1-g}\\
i \bar\partial q_{1} = p_{1} q_{0},
\label{q1-g}\\
i \bar\partial r_{0} = - p_{0} r_{1}. \label{r0-g}
\end{gather}
This system is in some sense intermediate between the DDEs
(\ref{1}), (\ref{-1}) and the PDE (\ref{o31}): both of them can be
`reconstructed' from (\ref{q0-g})--(\ref{r0-g}) (we will return to
this question below). And namely system (\ref{q0-g})--(\ref{r0-g})
is, strictly speaking, gauge equivalent to the spin field equation
we are dealing with.

Now we will derive the ZCR for the stationary $(2+1)$-dimensional
Landau--Lifshitz equation from (\ref{zcr-evol-g}) using the gauge
transformation by means of the matrix $\Psi(\lambda)$. Introducing the
matrix function $\Phi(\zeta)$
\begin{equation}
\Phi(\zeta) =  \Psi^{-1}(\lambda) \Psi( \zeta)
\label{phi-g}
\end{equation}
one can obtain from (\ref{zcr-evol-g}) that it satisfies the following
equations
\begin{equation}
\partial \Phi(\zeta) = V_{L}(\zeta) \Phi(\zeta)
,\qquad
\bar\partial \Phi(\zeta) = W_{L}(\zeta) \Phi(\zeta),
\label{zcr-evol-ll}
\end{equation}
where
\begin{gather}
V_{L}(\zeta)  =
  \Psi^{-1}(\lambda)
  \left[ V(\zeta) - V(\lambda) \right]
  \Psi(\lambda),
\\
  W_{L}(\zeta)  =
  \Psi^{-1}(\lambda)
  \left[ W(\zeta) - W(\lambda) \right]
  \Psi(\lambda).
\end{gather}
Noting that
\begin{gather}
S_{z}=
  \Psi^{-1}(\lambda)
  \left[ \sigma^{3}, V(\lambda) \right]
  \Psi(\lambda)
=
  2i \lambda^{-1}
  \Psi^{-1}(\lambda)
  \left(\begin{array}{cc} 0 & -r_{0} \\ q_{1} & 0 \end{array}\right)
  \Psi(\lambda),
\\
S_{\bar z}=
  \Psi^{-1}(\lambda)
  \left[ \sigma^{3}, W(\lambda) \right]
  \Psi(\lambda)
=
  2i \lambda
  \Psi^{-1}(\lambda)
  \left(\begin{array}{cc} 0 & r_{1} \\ -q_{0} & 0 \end{array}\right)
  \Psi(\lambda)
\end{gather}
one can present $V_{L}$ and $W_{L}$ as
\begin{gather}
V_{L}(\zeta) =
  \frac{ i}{2 } \left( \zeta^{-2} - \lambda^{-2} \right)
    \left( S - 1 \right) +
 \frac 12 \left( \lambda\zeta^{-1} - 1 \right)
    S S_{z},
\label{V-ll}
\\
W_{L}(\zeta) =
 \frac{ i}{2 } \left( \zeta^{2} - \lambda^{2} \right)
  \left( S + 1 \right) +
\frac 12 \left( \zeta\lambda^{-1} - 1 \right)
  S S_{\bar z}.
\label{W-ll}
\end{gather}
Using the zero-curvature conditions for equations~(\ref{zcr-evol-ll}),
\begin{equation}
\bar\partial V_{L} - \partial W_{L} + \left[ V_{L}, W_{L} \right]
= 0,
\end{equation}
and calculating the left-hand-side part of this equation
\begin{gather}
\bar\partial V_{L} - \partial W_{L} +
\left[ V_{L}, W_{L} \right] =
\nonumber\\
\qquad{}=
\frac{1}{2}
\left( \frac{\lambda}{\zeta} - \frac{\zeta}{\lambda} \right)
\left\{
  S S_{z \bar z} +
  \frac{1}{2} \left( S_{z} S_{\bar z} + S_{\bar z} S_{z} \right) +
  i \lambda^{2} S_{z} + i \lambda^{-2} S_{\bar z}
\right\}
\end{gather}
one can conclude that the matrix $S$ must solve
\begin{equation}
  S S_{z \bar z} +
  \frac{1}{2} \left( S_{z} S_{\bar z} + S_{\bar z} S_{z} \right) +
  i \lambda^{2} S_{z} + i \lambda^{-2} S_{\bar z} = 0.
\end{equation}
Noting that the anticommutator of traceless $2 \times 2$ matrices is
proportional to the unit one and that the anticommutator of $S$ and
$S_{z}$ or $S_{\bar z}$ is zero (which follows from the fact that
$\det S = 1$, which is another form of the equality
$\vec S^{2} = 1$) one can present this equation in the form
(\ref{main_eq}). Thus the linear problems (\ref{zcr-evol-ll}) together
with definitions (\ref{V-ll}) and~(\ref{W-ll}) can be viewed as the
ZCR for the main equation of the present paper.

After we have derived the ZCR for (\ref{main_eq}) we would like to
make a few remarks on the application of the inverse scattering
technique to non-evolutionary equations as ours. The IST has been
originally developed for the Cauchy problems. However, since then
much efforts has been made to adjust this method for various
boundary value problems. This is a rather difficult task since the
latter seem to be more difficult than the former ones. Among
successes in this field one should mention results related to the
hyperbolic systems such as, e.g, the sine-Gordon equation, the
principal chiral field equations etc. For these models the initial
value -- boundary value problems has been shown to be well stated
problems, the existence and uniqueness of the solution has been
established and IST-based algorithms to solve, say, the Goursat
type boundary problem have been elaborated. One can find
discussion of some recent results on boundary problems for the
$(1+1)$-dimensional systems on semi-infinite and finite interval,
for example in \cite{LeonSpire}.

As to the elliptical systems, similar to the one discussed here,
which do not possess characteristics, it is also possible to apply
the IST for solving some boundary value problems. Usually it is
achieved by breaking the symmetry between the coordinates (which
is, of course, not very natural for this kinds of equations),
selecting one of them (say, $y$), considering the problem on a
half-plane ($y>0$) or on a finite domain (in this case the part of
the boundary data plays role of the Cauchy conditions) and
performing analysis (i.e.\ solving the direct and inverse spectral
problems) for the auxiliary linear equation corresponding to the
complementary coordinate (say, $\Psi_{x} = U \Psi$). One can find
examples of such approach in \cite{GLN} and references therein.
However, in this paper we do not discuss the mathematically
rigorous formulation of the problem related to (\ref{veczz}). We
consider here the IST as a method to generate some classes of
particular solutions and restrict ourselves to the ones most
interesting from the physical viewpoint, solitons and
quasiperiodic solutions.

Above we have mapped the $V$-$W$ pair for system
(\ref{q0-g})--(\ref{r0-g}) into the $V_{L}$-$W_{L}$ pair for
equation (\ref{main_eq}) by means of the gauge transformation
(\ref{phi-g}),
\begin{gather}
V_{L}(\zeta) =
  - \Psi^{-1}(\lambda)  \partial \Psi(\lambda)
  + \Psi^{-1}(\lambda) V(\zeta) \Psi(\lambda),
\\
W_{L}(\zeta) =
  - \Psi^{-1}(\lambda)  \bar\partial \Psi(\lambda)
  + \Psi^{-1}(\lambda) \widetilde V(\zeta) \Psi(\lambda).
\end{gather}
Now we are going to derive the inverse transform: from (\ref{zcr-evol-ll})
to (\ref{zcr-evol-g}) (i.e.\ from the $V$-$W$ pair (\ref{V-ll}),
(\ref{W-ll}) to (\ref{V-alh}), (\ref{W-alh})). The fist step is to
diagonalize a solution of the Landau--Lifshitz equation, i.e., to
calculate, for given $S$, the matrix $\Psi$ defined by
\begin{equation}
S = \Psi^{-1} \sigma^{3} \Psi. \label{diag}
\end{equation}
It is obvious that the $S \rightarrow \Psi$ correspondence is not
one-to-one. For any $\Psi$ satisfying (\ref{diag}) the matrix
$D\Psi$ with an arbitrary diagonal matrix $D$ will also solve
(\ref{diag}). The main point of the Landau--Lifshitz equation
$\rightarrow$ ALH transform is to use this arbitrariness to
present the matrices $\partial\Psi \cdot \Psi^{-1}$,
$\bar\partial\Psi \cdot \Psi^{-1}$ in (\ref{V-alh}), (\ref{W-alh})
form with $\zeta=\lambda$.
\begin{equation}
\partial\Psi \cdot \Psi^{-1} = V(\lambda)
\qquad \mbox{and} \qquad
\bar\partial\Psi \cdot \Psi^{-1} = W(\lambda).
\end{equation}
This step needs some calculations which are presented in the
Appendix. Performing then the gauge transform with the found
matrix $\Psi$, one can obtain that the transformed $V_{L}$,
$W_{L}$ matrices
\begin{gather}
V(\zeta) =
  \partial\Psi \cdot \Psi^{-1} +
  \Psi V_{L}(\zeta) \Psi^{-1},
\\
W(\zeta) =
  \bar\partial\Psi \cdot \Psi^{-1} +
  \Psi \widetilde V_{L}(\zeta) \Psi^{-1}
\end{gather}
are exactly of the form (\ref{V-alh}), (\ref{W-alh}) which means that
the functions $q_{0}$, $r_{0}$, $q_{1}$, $r_{1}$ (which are defined now
in the terms of the matrix $\Psi$ (i.e.\ in the terms of the matrix $S$)
solve the system (\ref{q0-g})--(\ref{r0-g}).

System (\ref{q0-g})--(\ref{r0-g}) that can be rewritten as the
DDEs from the ALH.  Indeed, starting from the quantities $q_{0}$,
$r_{0}$, $q_{1}$, $r_{1}$ one can {\it define} the quantities
$q_{2}$, $r_{2}$
\begin{equation}
q_{2}  = i \frac{\partial q_{1}}{1 - q_{1}r_{1} }
,\qquad
r_{2}  = -i \frac{\bar\partial r_{1}}{1 - q_{1}r_{1} }
\end{equation}
and demonstrate that they satisfy the following identities:
\begin{equation}
i \bar\partial q_{2} = \left( 1 - q_{2}r_{2} \right) q_{1}
,\qquad
-i \partial r_{2} = \left( 1 - q_{2}r_{2} \right) r_{1}.
\end{equation}
Analogously, the quantities $q_{-1}$, $r_{-1}$,
\begin{equation}
q_{-1}  = i \frac{\bar\partial q_{0}}{1 - q_{0}r_{0} }
,\qquad
r_{-1}  = -i \frac{\partial r_{0}}{1 - q_{0}r_{0} }
\end{equation}
satisfy
\begin{equation}
i \partial q_{-1} = \left( 1 - q_{-1}r_{-1} \right) q_{0}
,\qquad
-i \bar\partial r_{-1} = \left( 1 - q_{-1}r_{-1} \right) r_{0}.
\end{equation}
This procedure can be repeated in both directions
\begin{equation}
\cdots
\leftarrow
( q_{-1}, r_{-1} )
\leftarrow
( q_{0}, r_{0}, q_{1}, r_{1} )
\rightarrow
( q_{2}, r_{2} )
\rightarrow
\cdots.
\end{equation}
This gives an infinite sequence of $q_{n}$'s, $r_{n}$'s which solve
\begin{gather}
i \partial q_{n} = p_{n} q_{n+1},
\\
-i \partial r_{n} = p_{n} r_{n-1}
\end{gather}
and
\begin{gather}
 i \bar\partial q_{n} = p_{n} q_{n-1},
\\
-i \bar\partial r_{n} = p_{n} r_{n+1},
\end{gather}
i.e.\ the Ablowitz--Ladik DDEs.

To conclude this section we want to discuss the following
question. If we start with the ALH, which is a system of DDEs,
then the relation between the {\it discrete} equations (ALH) and
the {\it partial differential} Landau--Lifshitz equation is rather
obvious: our PDE is a differential consequence of the DDEs. But if
we start with the Landau--Lifshitz equation, then what role do the
DDEs from the ALH play in the theory of our PDE? In simpler words,
what does the subscript $n$ mean in terms of our PDE? The answer
is as follows. We have an example of the situation studied by
Levi, Benguria \cite{LB,L}, Shabat, Yamilov \cite{SY} and others:
discrete integrable equations (the equations from the ALH in our
case) describe sequences of the B\"acklund transformations for
some PDEs (the stationary $(2+1)$-dimensional Landau--Lifshitz
equation in our case). Indeed, if we have a solution of our
equation, $\vec S_{1}$, we can derive from it the matrix
$\Psi_{1}$ which solves the linear problems for the DNLSE and
DMKdV, and hence the quantities $q_{0}$, $r_{0}$, $q_{1}$, $r_{1}$
which solve (\ref{q0-g})--(\ref{r0-g}). Then we can construct the
new $\Psi$-matrix $\Psi_{2}=\left(\begin{array}{cc} \lambda &
r_{1} \\ q_{1} & \lambda^{-1}\end{array}\right) \Psi_{1}$, and the
new spin field $\vec S_{2}$ which corresponds to the matrix
$\Psi_{2}^{-1} \sigma^{3} \Psi_{2}$. This vector field will also
solve the Landau--Lifshitz equation. This procedure can be
repeated infinitely
\begin{equation}
\cdots
\rightarrow
\vec S_{n}
\rightarrow
\Psi_{n}, q_{n}, r_{n}
\rightarrow
\Psi_{n+1} = U_{n}\Psi_{n}
\rightarrow
\vec S_{n+1}
\rightarrow
\cdots.
\end{equation}
Moreover, it can be performed in other direction
\begin{equation}
\cdots
\rightarrow
\vec S_{n}
\rightarrow
\Psi_{n}, q_{n}, r_{n}
\rightarrow
\Psi_{n-1} = U_{n-1}^{-1}\Psi_{n}
\rightarrow
\vec S_{n-1}
\rightarrow
\cdots.
\end{equation}
Thus we can obtain an infinite number of
solutions
( \ldots,
$\vec S_{-1}$, $\vec S_{0}$, $\vec S_{1}=\vec S$, $\vec S_{2}$,
\ldots)
from one solution $\vec S$
and relations between the vectors $\vec S_{n}$ with different values of
the index $n$ (B\"acklund relations) can be described by the equations which
are analogous to (and can be derived from) the DDEs from the ALH.

\section{Soliton structures} \label{sec_sol}

The discrete nonlinear Schr\"odinger equation (\ref{dnlse}) under
the condition (\ref{kappa}) has been already solved in the
pioneering work by Ablowitz and Ladik~\cite{AL2}. As to the
solutions of~(\ref{dmkdv}), or system (\ref{1}), (\ref{-1}), they
can be obtained by minor modifications of the ones for
(\ref{dnlse}), which is again a manifestation of the fact that all
of them belong to the same hierarchy. We will not repeat here the
derivation of the IST (one can find the technical details
in~\cite{AL2} or, say, in the book~\cite{AS}) and write down only
some final formulae that will be used below.

The $N$-soliton solution of equations (\ref{1}), (\ref{-1}) can be
presented as follows:
\begin{equation}
\bar q_{n}(z,\bar z) =
\sum_{j=1}^{N}
C_{nj}(z,\bar z) \xi_{nj} (z, \bar z),
\qquad
\xi_{nj} (z, \bar z) = \sum_{k=1}^{N} \left[ M^{-1}_{n}(z,\bar z)
\right]^{(jk)} \lambda_{k}^{-1}.
\end{equation}
The constants $\lambda_{k}$'s are the eigenvalues of the corresponding
scattering problem (\ref{zcr-sp}) (to be more precise, the discrete
spectrum of the scattering problem (\ref{zcr-sp}) consists of $N$ pairs of
the eigenvalues $(\lambda_{k}, -\lambda_{k})$). The functions
$C_{nk}(z,\bar z)$ are given by
\begin{equation}
C_{nk}(z,\bar z) =
C_{k}^{0} \lambda_{k}^{2n} \exp\left\{i \phi_{k}(z, \bar z) \right\},
\label{cnk}
\end{equation}
where $C_{k}^{0}$'s are arbitrary constants,
\begin{equation}
\phi_{k}(z,\bar z) =
\lambda_{k}^{-2} z + \lambda_{k}^{2} \bar z,
\end{equation}
while the matrix $M$ is given by
\begin{equation}
M = I + \Lambda \bar A_{n} \bar\Lambda A_{n}.
\end{equation}
Here $I$ is the $N \times N$ unit matrix, the overbar stands for the
complex conjugation,
\begin{equation}
\Lambda = {\rm diag} \,
\left( \lambda_{1} , \ldots , \lambda_{N} \right)
\end{equation}
and $A_{n}$ is the $N\times N$ matrix with the elements
\begin{equation}
A_{n}^{(jk)}(z,\bar z)  =
\frac{ C_{nk}(z,\bar z)}{1 - \bar\lambda_{j}^{2}\lambda_{k}^{2} }
\label{anjk}
\end{equation}
Solution for system (\ref{zcr-sp})--(\ref{zcr-evoln}) can be
presented in the pure soliton case as
\begin{equation}
\Psi_{n}(z,\bar z) =
F_{n}(\lambda; z,\bar z)
\left(\begin{array}{cc}
\exp( i \lambda^{2} \bar z ) & 0 \vspace{1mm}\\
0 & \exp( -i \lambda^{-2} z ) \end{array}\right),
\label{PsiF}
\end{equation}
where $F_{n}(\lambda)$ is the matrix of the following structure:
\begin{equation}
F_{n}(\lambda) =
\frac{ 1}{f_{n}(\infty) }
\left(\begin{array}{cc}
f_{n}(\lambda)
&
g_{n}(\lambda)
\vspace{2mm}\\
- \overline{ g_{n}(1/\bar\lambda)}
& \overline{ f_{n}(1/\bar\lambda)}\end{array}\right). \label{Fmatr}
\end{equation}
Here
\begin{equation}
f_{n}(\lambda) =
1 - \lambda^{2} \sum_{j,k=1}^{N}
\frac{ \bar C_{nj} \bar\lambda_{j} A_{n}^{(jk)} \xi_{nk}}
{1 - \lambda^{2}\bar\lambda_{j}^{2} }
, \qquad
g_{n}(\lambda) =
\lambda \sum_{j=1}^{N}
\frac{ C_{nj} \xi_{nj}}
{ \lambda^{2} - \lambda_{j}^{2} }.
\label{Psifg}
\end{equation}
The above formulae contain all we need to construct solutions for the
Landau--Lifshitz equation (\ref{veczz}), or (\ref{main_eq}). The vertical
component of the vector $\vec S_{n}$ (see (\ref{vecSn123})) can be
presented, using (\ref{PsiF})--(\ref{Psifg}) and the fact that in our case
$\lambda = \exp(i\gamma/2) = 1/\bar\lambda$ (see remark after
(\ref{veczz})), as
\begin{equation}
S_{n3} =
1 -
\frac{ 2 \left| g_{n}(\lambda) \right|^{2}}
{ f_{n}(\infty)}
\end{equation}
while the horizontal components can be written as
\begin{equation}
S_{n1} = S_{n\bot}\cos \varphi_{n},
\qquad
S_{n2} = S_{n\bot}\sin \varphi_{n},
\end{equation}
where
\begin{equation}
S_{n\bot} = \sqrt{1 - S_{n3}^{2}},
\qquad
\varphi_{n} =
  \arg { f_{n}(\lambda) } - \arg { g_{n}(\lambda) } +
  \lambda^{-2} z + \lambda^{2}\bar z.
\end{equation}
The magnetic energy density (\ref{W}) in this case can be presented,
using (\ref{Wqr}) and the identity $p_{n} = \det F_{n+1} / \det F_{n}$ as
\begin{equation}
{\cal W}[\vec S_{n}] = \frac{ v^{2}}{4 g} \left\{ \frac{ f_{n}(\infty) }
{f_{n+1}(\infty) } + \frac{ f_{n-1}(\infty)}{f_{n}(\infty) }
- 2 \right\}.
\end{equation}
Noting that, for a fixed value of the index $n$, the dependence on
$n$ can be taken into account by the redefinition of the constants
$C_{k}^{0}$, we may chose $n=0$ and write the final formulae as
follows:
\begin{equation}
  S_{3}(z, \bar z)  =
  1 -
  \frac{ 2 \left| g_{0}\left({\rm e}^{i\gamma/2}; z, \bar z\right) \right|^{2}}
{ f_{0}(\infty; z, \bar z) }
\label{S3}
\end{equation}
and
\begin{gather}
  S_{1}(z, \bar z)  =   S_{\bot}(z, \bar z)\cos \varphi(z, \bar z),
\label{S1}
\\
  S_{2}(z, \bar z)   =   S_{\bot}(z, \bar z)\sin \varphi(z, \bar z),
\label{S2}
\end{gather}
where
\begin{gather}
S_{\bot}(z, \bar z) = \sqrt{1 - S_{3}^{2}(z, \bar z)},
\\
\varphi(z, \bar z) =
\arg { f_{0}\left({\rm e}^{i\gamma/2}; z, \bar z\right) } -
\arg { g_{0}\left({\rm e}^{i\gamma/2}; z, \bar z\right) } +
{\rm e}^{-i\gamma} z + {\rm e}^{i\gamma} \bar z
\end{gather}
while the distribution of the magnetic energy of the field
configuration given by (\ref{S3})--(\ref{S2}) can be written as
\begin{equation}
  {\cal W} =
  \frac{ v^{2}}{4 g}
  \left\{
\frac{ f_{0}(\infty, z, \bar z)}{ f_{1}(\infty, z, \bar z) } +
\frac{ f_{-1}(\infty, z, \bar z)}{f_{0}(\infty, z, \bar z)} - 2
  \right\}.
\label{energy}
\end{equation}
These formulae describe the $N$-soliton solutions of the
stationary $(2+1)$-dimensional Lan\-dau--Lifshitz equation.

To make clear what kind of solutions we have obtained from the
ALH-solitons let us consider in a more detailed way the simplest of the
above solutions, namely the one-soliton ones. In this case the
quantity $A_{n}(z, \bar z) = A_{n}^{(11)}(z,\bar z)$ (see (\ref{anjk}))
using the designation
\begin{equation}
\lambda_{1}^{2} = \exp( - 2\delta + i\gamma_{1} )
\end{equation}
can be rewritten as
\begin{equation}
  A_{n}(z,\bar z) = \exp\left\{ \chi_{n} + i \psi_{n} \right\},
\end{equation}
where
\begin{gather}
\chi_{n}  =
  2 \sinh 2\delta
  \left(
  \sin\gamma_{1} \cdot {\rm Re}\, z -
  \cos\gamma_{1} \cdot {\rm Im}\, z
  \right) -
  2n \delta + \chi_{*},
\\
\psi_{n}  =
  2 \cosh 2\delta
  \left(
  \cos\gamma_{1} \cdot {\rm Re}\, z +
  \sin\gamma_{1} \cdot {\rm Im}\, z
  \right) +
  n \gamma_{1} + \psi_{*}
\end{gather}
and $\chi_{*}$, $\psi_{*}$ are some constants. Setting $n=0$,
returning to the real coordinates $x$, $y$ and~$t$, and
introducing the vectors
\begin{equation}
\vec k_{\bot} =
  \frac{ v \sinh 2\delta}{2 g }
  \left(\begin{array}{c} \sin\gamma_{1} \\ - \cos\gamma_{1} \end{array}\right),
\qquad
\vec k_{\|} =
  \frac{ v \cosh 2\delta}{2 g }
  \left(\begin{array}{cc} \cos\gamma_{1} \\ \sin\gamma_{1} \end{array}\right),
\end{equation}
one can rewrite these formulae as
\begin{equation}
  A(x,y,t) =  \exp\left\{ \chi(x,y,t)+i\psi(x,y,t) \right\}
\label{A1}
\end{equation}
with
\begin{gather}
\chi(x,y,t) =
  \left( \vec k_{\bot} , \vec r - \vec v t \right) + \mbox{const},
\label{chi1}\\
\psi(x,y,t) =
  \left( \vec k_{\|} , \vec r - \vec v t \right) + \mbox{const}
\label{psi1}
\end{gather}
(here braces stand for the usual scalar product: for
$\vec k = {\mathrm{col}} ( k_{x}, k_{y})$,
$( \vec k, \vec r ) = k_{x}x + k_{y}y$),
from which one can derive the following expressions for the components of
the vector $\vec S$.
The vertical components, $S_{3}$, can be written as
\begin{equation}
S_{3} =
1 -
\frac{ 2\sinh^{2} 2\delta}{ \cosh 2\delta -\cos 2\Gamma} \;
\frac{ 1}{\cosh 2\chi + \cosh 2\delta}
\label{S3one}
\end{equation}
where $\Gamma = (\gamma_{1} - \gamma)/2$, or, equivalently, as
\begin{equation}
S_{3} = \cos\theta
\label{theta}
\end{equation}
with the angle $\theta$ being given by
\begin{equation}
\tan \frac{\theta}{2} =
\left[
\left( \frac{ \sin \Gamma}{\sinh \delta } \cosh \chi \right)^{2} +
\left(\frac{ \cos \Gamma}{\cosh \delta } \sinh \chi \right)^{2}
\right]^{-1/2}.
\label{sol_tan}
\end{equation}
The horizontal components $S_{1,2}$ can be presented as
\begin{equation}
\left(\begin{array}{c} S_{1} \\ S_{2} \end{array}\right) =
\sin \theta
\left(\begin{array}{c} \cos\varphi \\ \sin\varphi\end{array}\right)
\label{varphi}
\end{equation}
with
\begin{equation}
\varphi(x,y,t) =
  \hat\varphi(x,y,t) +
  \left( \vec k_{0} - \vec k_{\|} , \vec r - \vec v t \right) +
  \mbox{const}.
\label{varphione}
\end{equation}
Here, the vector $\vec k_{0}$,
\begin{equation}
\vec k_{0} =
  \frac{ v }{2 g }
  \left(\begin{array}{c} \cos\gamma \\ \sin\gamma \end{array}\right),
\end{equation}
is parallel to the velocity vector, $\vec k_{0} = \vec v / 2g$,
and is related to $\vec k_{\bot}$, $\vec k_{\|}$ by $\vec
k_{0}^{2} = \vec k_{\|}^{2} - \vec k_{\bot}^{2}$. The function
$\hat\varphi$ is given by
\begin{equation}
\hat\varphi = \arctan \left( \tanh\delta \, \cot\Gamma \,
\tanh\chi \right).
\end{equation}

The magnetic energy of this field configuration can be written as
\begin{equation}
{\cal W} = 4 g k_{\bot}^{2}
\frac{ 1 + \cosh 2\delta \, \cosh 2\chi}
{\left( \cosh 2\chi + \cosh 2\delta \right)^{2} },
\label{energy_1}
\end{equation}
where $k_{\bot}$ stands for $|\vec k_{\bot}|$.

It can be shown that ${\cal W}$ (\ref{energy_1}) is a second-order
polynomial in $S_{3}$:
\begin{equation}
{\cal W} =
- g \vec k_{\|}^{2} S_{3}^{2} +
  \left( \vec v, \vec k_{\|} \right) S_{3} +
  {\cal W}_{0},
\end{equation}
where ${\cal W}_{0}$ is some constant.
The linear energy density, ${\cal W}_{\rm lin}$,
\begin{equation}
{\cal W}_{\rm lin} =
\int_{-\infty}^{\infty}
d x_{\bot}
{\cal W} (x,y,t),
\qquad
x_{\bot} =
\left( \frac{\vec k_{\bot}}{k_{\bot}} \, , \vec r \right)
\end{equation}
can be easily shown to be
\begin{equation}
{\cal W}_{\rm lin} = 4 g k_{\bot}.
\end{equation}

To simplify the following analysis let us consider the case when
the velocity vector is directed along the $x$-axis ($\gamma=0$,
$\lambda=1$). This does not lead to loss of generality because
solutions corresponding the arbitrary vector $\vec v = (
v\cos\gamma, v\sin\gamma )$ can be obtained from the ones
presented below by the substitution $x \rightarrow x\cos\gamma +
y\sin\gamma$, $y \rightarrow y\cos\gamma - x\sin\gamma$.  It can
be easily seen that formulae (\ref{S3one})--(\ref{varphione}) in
the limiting cases $\gamma_{1}=\pi/2$ and $\gamma_{1}=0$ describe
essentially different field structures. In the case
$\gamma_{1}=\pi/2$ ($\Gamma=\pi/4$)
\begin{equation}
\chi = \chi(x-vt) = k_{\bot}\left( x - x_{*} - vt \right)
\end{equation}
with an arbitrary constant $x_{*}$, and both $S_{3}$ and ${\cal W}$ depend
on $x - v t$ only,
\begin{gather}
\theta = \theta( x - v t ) =
2 \arctan \left(
\frac{ \sinh 2\delta}
{  \left[
  \cosh^{2} ( \chi + \delta ) + \cosh^{2} ( \chi - \delta )
  \right]^{1/2}
} \right),\\
{\cal W} = {\cal W}( x - v t ) =
g k_{\|}^{2} \sin^{2} \theta( x - v t )
\end{gather}
while $\varphi$ can be written as
\begin{equation}
  \varphi(x,y,t) =
  \hat\varphi(x-vt) + k_{0} (x - vt) + k_{\|} y + \mbox{const},
\end{equation}
where
\begin{equation}
\hat\varphi(x) = \arctan \left( \tanh\delta\, \tanh
k_{\bot}(x-x_{*}) \right).
\end{equation}
So, this solution describes a localized structure, moving in the
$x$-direction, which is phase modulated in the transversal direction
($y$-direction), and it may be termed `quasi one-dimensional soliton' (see
Fig.~1).

\begin{figure}[t]
\centerline{\includegraphics[width=14cm]{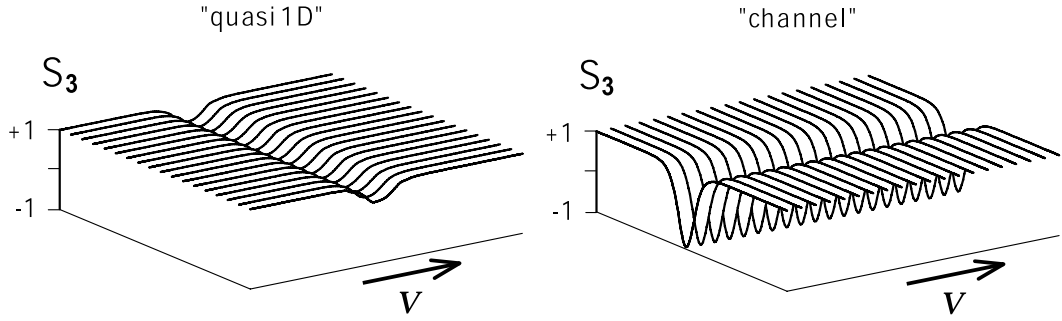}}
\caption{ Limiting cases of the one-soliton solutions (schematically) corresponding
  to $\Gamma=\pi/4$ and $\Gamma=0$.}
\end{figure}

The soliton obtained above is essentially two-dimensional
structure and despite apparent similarity it cannot be reduced to
its one-dimensional analogue. Indeed, in the one-dimensional case
soliton solutions of equation (\ref{heis11}) possess the following
form:
\begin{equation}
\theta = \theta\left(\frac{x - vt}{L(\Omega)} \right),
\qquad
\varphi = \Omega t + \hat\varphi\left(\frac{x - vt}{L(\Omega)} \right),
\end{equation}
and one can say that such solitons exist due to the {\it temporal}
phase modulation of the whole medium, which manifests itself in
the fact that $L(\Omega) \propto \Omega^{-1/2}$, i.e., soliton
vanishes with~$\Omega$ going to zero. In other words, these
one-dimensional soliton structures do not exist in absence of the
phase modulation $\Omega t$. In our case, existence of solitons is
due to the {\it spatial} phase modulation (in $y$-direction),
which manifests itself in the fact that the magnetic energy
density is proportional to $\vec k_{\|}^{2}$. In other words,
solitons we have obtained differ from their one-dimensional
analogues in the physical mechanism lying in their background:
they are caused by the competition between the {\it spatial}
dispersion and nonlinearity while in the one-dimensional case
solitons are caused by the competition between the {\it temporal}
dispersion and nonlinearity.

In the opposite case, $\gamma_{1} = 0$ ($\Gamma = 0$),
\begin{equation}
\theta = \theta( y ) = 2 \arctan \frac{ \cosh \delta}{\sinh
k_{\bot} ( y - y_{*} ) },
\end{equation}
i.e.\ $S_{3}$, which can be written as
\begin{equation}
S_{3} = S_{3}(y) =
\frac{  \sinh^{2} k_{\bot}\left( y-y_{*} \right) -
  \cosh^{2}\left( \delta_{1} / 2 \right)}
{ \sinh^{2} k_{\bot}\left( y-y_{*} \right) +
  \cosh^{2}\left( \delta_{1} / 2 \right)},
\end{equation}
depends on $y$ only (and, what is essential, does not depend on
time), while the horizontal components are rotating with constant
frequency:
\begin{equation}
\varphi = \varphi(x-vt) = \kappa_{x}(x-vt),
\end{equation}
where
\begin{equation}
\kappa_{x} = k_{\|} - k_{0}
\end{equation}
and $k_{a}$, remind, stands for $|\vec k_{a}|$.

This solution describes the spin wave localized in the
$2k_{\bot}^{-1}$-neighborhood of the line $y=y_{*}$ (this field
distribution, which is depicted schematically in the Fig.~1, may be termed
`channel').  The magnetic energy of the `channel' field configuration does
not depend on time, ${\cal W}={\cal W}(y)$, hence it can be considered as
almost static, in the sense that we have no energy transport in this case.
Similar structures have been found by A~S~Kovalev~\cite{K}.

\begin{figure}[th]
\centerline{\includegraphics[width=10cm]{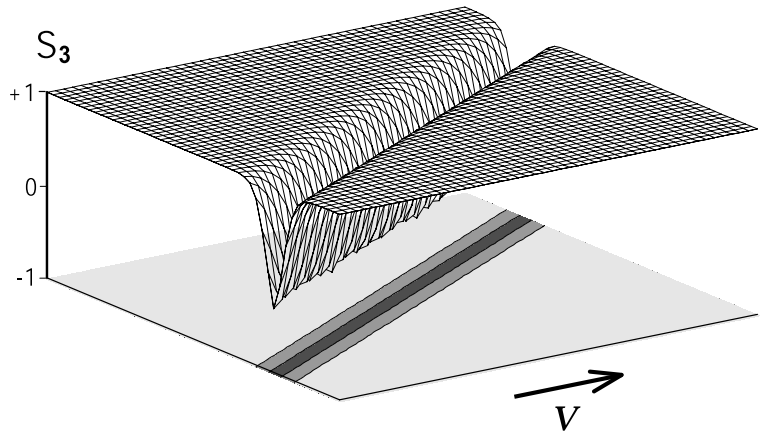}}
\caption{One-soliton solution for $\lambda=1$ and $\lambda_1=0.8\exp(i\pi/12)$ }
\end{figure}

\begin{figure}[th]
\centerline{\includegraphics[width=10cm]{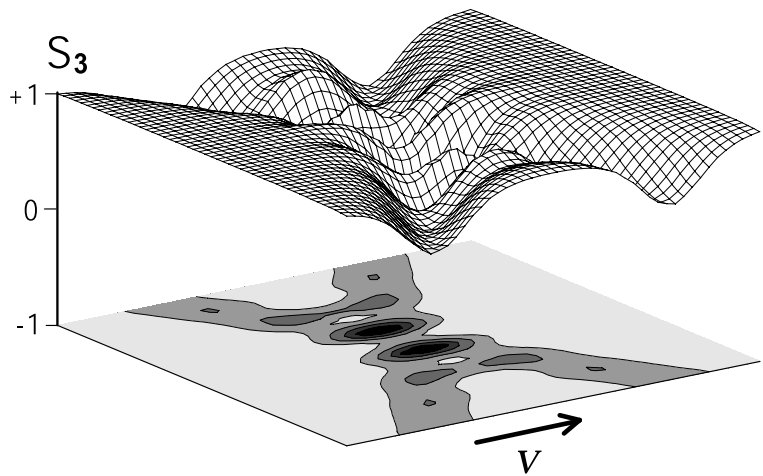}}
\caption{Two-soliton solution for $\lambda=1$ and
  $\lambda_1=\bar\lambda_2=0.8\exp(i\pi/12)$}
\end{figure}

The character of the soliton field structure in the general case
$0 < \gamma_{1} < \pi/2$ can be seen from the Fig.~2. The
many-soliton structures in the general case can be viewed as
consisting of several intersecting solitons. One can find typical
two-soliton spin distribution in the Fig.~3.

\section{Quasiperiodic structures} \label{sec_QPS}

The ALH in the quasiperiodic case is less studied than in the
soliton one. Several authors have discussed the quasiperiodic
solutions (QPS) for the discrete nonlinear Schr\"odinger and the
discrete modified Korteveg-de Vries equations (see, e.g.,
\cite{Bog,AC}), but their results are not enough to construct the
corresponding solutions for the Heisenberg equation using the
`embedding into the ALH' method. What we need and what is absent
in the papers~\cite{Bog,AC} is a solution of the auxiliary system
(\ref{zcr-sp})--(\ref{zcr-evoln}) which in the quasiperiodic case
is known as the Baker--Akhiezer function. Later this question was
solved in \cite{MEKL} (see also~\cite{QPS}). However, these
results, describing general finite-genus solutions, are rather
cumbersome, so here we restrict ourselves only to the elliptic
solutions, which are the simplest QPS.

The elliptic solutions for system (\ref{1}), (\ref{-1}) possess
the following structure:
\begin{equation}
q_{n} = q_{*} \varepsilon^{n} {\rm e}^{i\phi}
  \frac{\vartheta_{1}(\zeta_{n}+\alpha)}{\vartheta_{1}(\zeta_{n}) },
\qquad
r_{n} = r_{*} \varepsilon^{-n} {\rm e}^{-i\phi}
\frac{\vartheta_{1}(\zeta_{n}-\alpha)}{\vartheta_{1}(\zeta_{n}) },
\label{qp_qr}
\end{equation}
where $\vartheta_{1}$ is one of the elliptic theta-functions (see,
e.g.~\cite{Bateman}), the phase $\phi$ is some linear function of the
coordinates $z$ and $\bar z$ (it will be specified below),
\begin{equation}
\zeta_{n} =
\frac{z}{L} + \frac{\bar z}{\tilde L} + n \beta + \mbox{const}
\end{equation}
and the constants $q_{*}$, $r_{*}$ are related by
\begin{equation}
q_{*}r_{*} =
- \frac{ \vartheta_{1}^{2}(\beta)}
{  \vartheta_{1}(\alpha+\beta)\vartheta_{1}(\alpha-\beta) }.
\end{equation}
The quantities $p_{n}$'s can be presented as
\begin{equation}
p_{n} =
\frac{ \vartheta_{1}^{2}(\alpha)}
{ \vartheta_{1}(\alpha+\beta)\vartheta_{1}(\alpha-\beta) }
\frac{ \vartheta_{1}(\zeta_{n}+\beta)\vartheta_{1}(\zeta_{n}-\beta)}
{ \vartheta_{1}^{2}(\zeta_{n}) }
\end{equation}
which follows from expressions (\ref{qp_qr}) and the identity
\begin{gather}
\vartheta_{1}^{2}(x)\vartheta_{1}(y+z)\vartheta_{1}(y-z) +
\vartheta_{1}^{2}(y)\vartheta_{1}(z+x)\vartheta_{1}(z-x) \nonumber\\
\qquad{}+
\vartheta_{1}^{2}(z)\vartheta_{1}(x+y)\vartheta_{1}(x-y) = 0.
\label{Fay}
\end{gather}
This identity is the Fay's formulae \cite{Mumford} for the
elliptic functions. It can be used to calculate the derivatives of
the $\vartheta$-functions. Differentiating (\ref{Fay}) with
respect to $z$ and putting $z=y$ one can obtain for the
logarithmic derivative $\psi$,
\begin{equation}
\psi(\zeta) = \frac{ {\rm d}}{{\rm d}\zeta } \ln
\vartheta_{1}(\zeta),
\end{equation}
the relation:
\begin{equation}
\psi(\zeta+x) - \psi(\zeta-x) =
2 \psi(x) -
\frac{ \vartheta_{1}'(0) \vartheta_{1}(2x)}
{  \vartheta_{1}^{2}(x)}
\frac{ \vartheta_{1}^{2}(\zeta)}
{ \vartheta_{1}(\zeta+x) \vartheta_{1}(\zeta-x)}.
\label{psi_Fay}
\end{equation}
Using the latter one can obtain that functions (\ref{qp_qr})
satisfy equations (\ref{1}), (\ref{-1}) provided the scales $L$
and $\tilde L$ are chosen as
\begin{equation}
L =
i \vartheta_{1}'(0)
\frac{ \vartheta_{1}(\alpha-\beta)}
{    \vartheta_{1}(\alpha) \vartheta_{1}(\beta) }
  \varepsilon^{-1},
\qquad
\tilde L =
-i \vartheta_{1}'(0)
\frac{ \vartheta_{1}(\alpha+\beta)}
{     \vartheta_{1}(\alpha) \vartheta_{1}(\beta) }
  \varepsilon,
\end{equation}
and the phase $\phi$ is given by
\begin{equation}
\phi =
- i \left[ \psi(\beta) - \psi(\beta+\alpha) \right]
  \frac{ z}{L }
+ i \left[ \psi(\beta) - \psi(\beta-\alpha) \right]
  \frac{ \bar z}{\tilde L }
+ \mbox{const}.
\end{equation}

The Baker--Akhiezer function of our problem (i.e.\ the quasiperiodic
solution for system~(\ref{zcr-evolp})), (\ref{zcr-evoln}) can be
written as a matrix with the elements
\begin{gather}
\Psi^{(11)}_{n}(\lambda) =
  A_{1}\mu_{1}^{n}
\frac{ \theta(\zeta_{n}+\eta)}{ \theta(\zeta_{n}-\beta) }
  \exp(i\Phi_{1}),
\\
\Psi^{(21)}_{n}(\lambda) =
  -A_{1}\mu_{1}^{n}
  \lambda^{-1} q_{n-1}
  \frac{ \theta(\alpha-\beta)\theta(\eta)}
{    \theta(\beta)\theta(\eta+\alpha) }
 \frac{ \theta(\zeta_{n}+\eta+\alpha)}
{\theta(\zeta_{n}+\alpha-\beta) }
  \exp(i\Phi_{1}),
\\
\Psi^{(12)}_{n}(\lambda) =
  A_{2}\mu_{2}^{n}
  \lambda r_{n-1}
  \frac{ \theta(\alpha+\beta)\theta(\eta+\beta)}
{    \theta(\beta)\theta(\eta+\alpha+\beta) }
 \frac{ \theta(\zeta_{n}-\eta-\alpha-\beta)}
{ \theta(\zeta_{n}-\alpha-\beta) }
  \exp(i\Phi_{2}),
\\
\Psi^{(22)}_{n}(\lambda) =
  A_{2}\mu_{2}^{n}
  \frac{ \theta(\zeta_{n}-\eta-\beta) }{\theta(\zeta_{n}-\beta) }
  \exp(i\Phi_{2}).
\end{gather}
Here $A_{1,2}$ are arbitrary constants which are of no importance for our
further consideration. The phases $\Phi_{1,2}$ are the linear functions
of the coordinates,
\begin{gather}
\Phi_{1}=
\left[ \psi(\eta + \alpha + \beta) - \psi(\alpha) \right]
  \frac{ i z}{L } +
\left[
  \psi(\eta+\beta) + \psi(\alpha) -
  \psi(\alpha-\beta) - \psi(\beta)
\right]
  \frac{ i \bar z}{\tilde L },
\\
\Phi_{2} =
\left[
  - \psi(\eta) + \psi(\alpha+\beta) -
  \psi(\alpha) - \psi(\beta)
\right]
  \frac{ i z}{L } +
\left[ - \psi(\eta + \alpha) + \psi(\alpha) \right]
  \frac{ i \bar z}{\tilde L },
\end{gather}
the quantities $\mu_{1,2}$ are given by
\begin{equation}
\mu_{1} =
  \lambda
  \frac{ \theta(\alpha)\theta(\eta+\alpha)}{ \theta(\alpha-\beta)\theta(\eta+\alpha+\beta) },
\qquad
\mu_{2} =
  \lambda^{-1}
\frac{ \theta(\alpha)\theta(\eta+\alpha+\beta)}
{\theta(\alpha+\beta)\theta(\eta+\alpha) }
\end{equation}
and $\eta$ can be determined as a solution of the equation
\begin{equation}
\frac{\vartheta_{1}(\alpha-\beta)}
{\vartheta_{1}(\alpha+\beta) } \,
\frac{\vartheta_{1}(\eta) \vartheta_{1}(\eta+\alpha+\beta)}
{\vartheta_{1}(\eta+\alpha) \vartheta_{1}(\eta+\beta) }
= \varepsilon \lambda^{2}
\label{eta}
\end{equation}
(in the framework of the general theory, $\eta$ can be considered as the
point of the Riemann surface that corresponds to the point
$\varepsilon\lambda^{2}$ of the complex plane).

These formulae (we do not present here the corresponding derivation
procedure) can be verified straightforwardly using (\ref{psi_Fay}) and
(\ref{Fay}). They provide all necessary to construct the elliptic
solutions for the Heisenberg equation (\ref{main_eq}).
Using (\ref{vecSn123}), (\ref{theta}) and (\ref{varphi}), and
omitting the $n$-dependence one can obtain
\begin{equation}
\tan^{2}\frac{\theta}{2} =
 -\frac{\vartheta_{1}(\eta) \vartheta_{1}(\eta+\beta)}
{   \vartheta_{1}(\eta+\alpha)\vartheta_{1}(\eta+\alpha+\beta) }
 \frac{\vartheta_{1}(\zeta+\eta+\alpha)\vartheta_{1}(\zeta-\eta-\alpha-\beta)}
{  \vartheta_{1}(\zeta+\eta)\vartheta_{1}(\zeta-\eta-\beta) }
\label{tg_theta_ell}
\end{equation}
and
\begin{equation}
\exp\left\{ 2i \varphi \right\} =
  \frac{\vartheta_{1}(\zeta+\eta)\vartheta_{1}(\zeta+\eta+\alpha)}
{   \vartheta_{1}(\zeta-\eta-\beta)\vartheta_{1}(\zeta-\eta-\alpha-\beta) }
  \exp\left\{ 2i\Phi \right\},
\end{equation}
where
\begin{equation}
\Phi = \Phi_{1} - \Phi_{2} + \phi.
\end{equation}
The last two formulae can be rewritten as
\begin{gather}
\varphi  =
  \widehat\varphi(\zeta) -\widehat\varphi^{'}(0)\zeta -
  \left( \vec \kappa_{\|}, \vec r - \vec v t \right),
\label{varphi_ell}
\\
\widehat\varphi(\zeta)  =
  \frac{1}{2i} \ln
  \frac{\vartheta_{1}(\zeta+\eta)\vartheta_{1}(\zeta+\eta+\alpha)}
{  \vartheta_{1}(\zeta-\eta-\beta)\vartheta_{1}(\zeta-\eta-\alpha-\beta) }.
\label{hatvarphi_ell}
\end{gather}
Here the vector $\vec \kappa_{\|}$ is given by
\begin{equation}
\vec \kappa_{\|} = \kappa_{\|}
  \left(\begin{array}{c} \cos\gamma_{1} \\ \sin\gamma_{1} \end{array}\right)
\label{kappa_par_ell}
\end{equation}
with
\begin{equation}
\kappa_{\|}^{2} =
  \frac{ v^{2}}{4 g^{2} } \,
  \frac{\vartheta_{1}^{4}(\alpha)\vartheta_{1}^{4}(\beta)
   \vartheta_{1}^{2}(2\eta+\alpha+\beta)}
{   \vartheta_{1}^{2}(\eta)\vartheta_{1}^{2}(\eta+\alpha)
   \vartheta_{1}^{2}(\eta+\beta) \vartheta_{1}^{2}(\eta+\alpha+\beta)
   \vartheta_{1}(\alpha+\beta) \vartheta_{1}(\alpha-\beta) }
\end{equation}
and the angle $\gamma_{1}$ is defined by
\begin{equation}
{\rm e}^{2i\gamma_{1}} =
  \frac{ 1}{ \varepsilon^{2} } \,
\frac { \vartheta_{1}(\alpha-\beta)}{\vartheta_{1}(\alpha+\beta) }.
\end{equation}

The magnetic energy density ${\cal W}$ (\ref{W}) of the above field
configuration is, as in the one-soliton case, a second-order polynomial in
$S_{3}$:
\begin{equation}
{\cal W} =
- g \vec \kappa_{\|}^{2} S_{3}^{2} +
  \left( \vec v, \vec \kappa_{\|} \right) S_{3} +
  {\cal W}_{0},
\end{equation}
where ${\cal W}_{0}$ is some constant.
The last formula again illustrates the importance of the transversal
modulation (space dispersion) for the existence of our nonlinear
structures.

It should be noted that to ensure reality of all physical
quantities, such as $S_{i}$, ${\cal W}$ one has to impose some
restrictions on the parameters $\alpha$, $\beta$ and $\eta$ (or
$\varepsilon$) which appear in the above expressions.  We cannot
at present formulate these restrictions in their general form, but
will show below how these parameters should be chosen in some
particular case, which is a~generalization of the pure soliton
one, in the sense that the one-soliton solutions obtained in
Section~\ref{sec_sol} are some limiting cases ot the elliptical
ones discussed below.

Thus, in what follows we restrict ourselves with the case of
\begin{equation}
\alpha = \frac{\tau}{2},
\end{equation}
where $\tau$ is the complex half-period of the $\vartheta_{1}$-function
(see~\cite{Bateman}).  It can be shown that in this case both the
components of the vector $\vec S$ and the energy ${\cal W}$ will be real
if we choose
\begin{equation}
\zeta = \frac{1+\beta}{2} - i\hat\zeta, \qquad
\eta  = \hat\eta - \frac{\beta}{2},  \qquad
\beta = 2i\hat\beta,
\end{equation}
where hats indicate that correspondent quantities are real. In
what follows we use together with the theta-functions also the
Jacobian elliptical functions ${\rm sn}$, ${\rm cn}$ and ${\rm
dn}$,
\begin{equation}
{\rm sn}\, u =
{\rm sn}\,(u,k) =
  \frac{\vartheta_{3}(0)}{\vartheta_{2}(0) } \,
\frac{ \vartheta_{1}(u/2K)}{\vartheta_{0}(u/2K) }
\end{equation}
with
\begin{equation}
k = \frac{ \vartheta_{2}^{2}(0)}{\vartheta_{3}^{2}(0) },
\qquad
K = \frac{\pi}{2} \vartheta_{3}^{2}(0)
\label{ell_k}
\end{equation}
(the definition of $\vartheta_{2,3}$ and analogous formulae for
${\rm cn}\, u$ and ${\rm dn}\, u$ one can find, e.g.,
in~\cite{Bateman}). The `coordinate' $\hat\zeta$ can be written now
as
\begin{equation}
\hat\zeta =
   \frac{ 1}{\pi }
   \left( \vec k_{\bot}, \vec r - \vec v t \right),
\end{equation}
where
\begin{equation}
\vec k_{\bot} =
  \frac{ v}{2g } \, \frac{ \pi}{2 K } \,
\frac { {\rm sn}\, 4iK\hat\beta}{i } \left(
  \begin{array}{c} \sin\gamma_{1} \\ - \cos\gamma_{1} \end{array}\right).
\end{equation}
The expressions for $\theta$ and $\varphi$
(\ref{tg_theta_ell}) and (\ref{varphi_ell}), (\ref{hatvarphi_ell}) can be
presented as
\begin{equation}
\tan^{2}\frac{\theta}{2 } =
  \left|
  {\rm sn}\left( 2K\hat\eta + 2iK\hat\beta \right) \,
  \frac{ {\rm dn}\left( 2K\hat\eta + 2iK\hat\zeta \right)}
{ {\rm cn}\left( 2K\hat\eta + 2iK\hat\zeta \right) }
  \right|^{2}
\label{ell_tan}
\end{equation}
and
\begin{equation}
\varphi  =
  \left( \vec \kappa_{0} - \vec \kappa_{\|}, \vec r - \vec v t \right) +
  {\rm arg} \,
  \left[
     \vartheta_{2}( \hat\eta - i\hat\zeta )
    \vartheta_{3}( \hat\eta - i\hat\zeta )
  \right].
\label{ell_varphi}
\end{equation}
The vector $\vec\kappa_{0}$ is given by
\begin{equation}
\vec \kappa_{0} =
  \frac{ 2}{\pi }
  \left[ {\rm Re}\, \psi_{0} \left(\hat\eta+i\hat\beta\right) \right]
  \vec k_{\bot},
\end{equation}
where
\begin{equation}
\psi_{0} (z) =
  \psi(z) -
  K \frac{ {\rm cn}\, 2Kz \; {\rm dn}\, 2Kz}{{\rm sn}\, 2Kz  } =
  \psi(z) -
  \frac{ \pi}{ 2 } \vartheta_{0}^{2}(0)
\frac{ \vartheta_{2}(z)\vartheta_{3}(z)}{\vartheta_{1}(z)\vartheta_{0}(z) }
\end{equation}
while the vector $\vec\kappa_{\|}$ is given by (\ref{kappa_par_ell}) with
\begin{equation}
\kappa_{\|} =
  i \, \frac{ v}{g } \,
  \frac{  {\rm sn}\, 4iK\hat\beta}{{\rm sn}\, 4K\hat\eta }
  \sin 2\Gamma
\end{equation}
and $\Gamma = (\gamma_{1} - \gamma) / 2$ is related to $\hat\beta$ and
$\hat\eta$ by
\begin{equation}
{\rm e}^{2 i \Gamma}  =
 \frac{ {\rm sn}\, 2K(\hat\eta + i\hat\beta)}
{{\rm sn}\, 2K(\hat\eta - i\hat\beta)}.
\end{equation}

It is straightforward to show that the limiting case of the
elliptic quasiperiodic solutions presented above is solitons
obtained in Section~\ref{sec_sol}. Indeed, with the parameter $k$
(\ref{ell_k}) going to zero (which corresponds to $\tau
\rightarrow i\infty$), the elliptic functions sn, cn and dn become
sin, cos and 1 correspondingly. Noting that $K(k=0)=\pi/2$ and
identifying $\pi\hat\beta$ with $\delta$ (which implies
$\pi\hat\zeta \rightarrow \chi$) one can transform (\ref{ell_tan})
and (\ref{ell_varphi}) to formulae (\ref{sol_tan}) and
(\ref{varphione}) describing solutions of the Landau--Lifshitz
equation in the one-soliton case.

\section{Conclusion}

To conclude, we want to summarize the main results and to outline
some perspectives of the studies discussed in this paper. From the
mathematical point of view, our main result is the established
relation between the Landau--Lifshitz equation (in the case $\vec
S(\vec r, t) = \vec S (\vec r - \vec v t)$) and the ALH. And
though we cannot at present provide general explanation of what
makes such apparently different models be so closely connected, we
hope that the results presented in Sections \ref{sec_sol},
\ref{sec_QPS} are rather convincing arguments in favour of the
fact that this relation is useful, at least from the practical
standpoint, as a tool for generating of a large number of
solutions. On the other hand, this work presents 2D stationary
structures of the isotropic continuous Heisenberg ferromagnetic
spin system which have not been, to our knowledge, discussed in
the literature and which seem to be interesting for the physics of
magnetic phenomena. It should be noted that we have obtained our
results in the framework of the classical model, and one of the
most important questions that should be solved now, from the
viewpoint of applications to magnetism, is to develop quantum, or
at least semi-classical, theory of such structures. Another
question we want to mention here is the following one. It is a
widely known fact that solitons appear as a result of joint action
of nonlinearity and some other mechanisms, such as dispersion. In
our consideration we have neglected the temporal dispersion
(temporal modulation), and its role has been played by the spatial
one. So, it is interesting to take into account both temporal and
spatial dispersions, because the competition of different
mechanisms in nonlinear regime can lead to nontrivial results.
These and some other related questions may be the subject of
further investigations.

\subsection*{Acknowledgements}

This work was partly carried out during the authors' stay at the
Abdus Salam International Centre for Theoretical Physics which is
gratefully acknowledged for its kind hospitality and was partly
supported by the Ministerio de Educaci\'{o}n, Cultura y Deporte of
Spain under grant SAB2000-0256. We are grateful to A~S~Kovalev
for useful and stimulating discussions.

\section*{Appendix}
\def\theequation{A.\arabic{equation}}
\setcounter{equation}{0}

The aim of this section is to derive the matrix $\Psi$ related to
the solution of equation (\ref{main_eq}) $S$ by
\begin{equation}
S = \Psi^{-1} \sigma^{3} \Psi \label{app:diag}
\end{equation}
such that the matrices $V$, $W$
\begin{equation}
V = \partial\Psi \cdot \Psi^{-1} ,\qquad W = \bar\partial\Psi
\cdot \Psi^{-1}
\end{equation}
have the structure of the ALH matrices (\ref{V-alh}),
(\ref{W-alh}). The diagonalization (\ref{app:diag}) of a given
matrix $S$ is not unique. Suppose we have found a matrix $F$
satisfying
\begin{equation}
S = F^{-1} \sigma^{3} F \label{app:diag-F}.
\end{equation}
Then any matrix
\begin{equation}
\Psi = D F \label{app:psi-df}
\end{equation}
with an arbitrary diagonal matrix $D=\mathrm{diag} \left( D_{11},
D_{22} \right)$ satisfies (\ref{app:diag}). Hence, to solve our
problem we can start from any solution of (\ref{app:diag-F}), for
example from one given by
\begin{equation}
F = 1 + \sigma^{3} S \label{app:F}
\end{equation}
(one can verify by simple calculations that this is indeed a
solution of (\ref{app:diag-F})) and then to construct the diagonal
matrix $D$ such that the matrices
\begin{gather}
V = \partial\Psi \cdot \Psi^{-1} =
   \partial D \cdot D^{-1} + D \, \partial F \cdot F^{-1} \, D^{-1},
\label{app:V-psi}
\\
W = \bar\partial\Psi \cdot \Psi^{-1} =
   \bar\partial D \cdot D^{-1} + D \, \bar\partial F \cdot F^{-1} \, D^{-1}
\label{app:W-psi}
\end{gather}
possess the properties we need.

To simplify the following formulae let us introduce the
designation $a$, $b$ and $c$ for the elements of the matrix $S$,
\begin{equation}
S = \left(\begin{array}{cc} a & b \\ c & -a \end{array}\right).
\end{equation}
The main equation of this paper (\ref{app:diag-F}) can be
rewritten now as a system
\begin{gather}
 a_{z \bar z} + a X + \frac{i}{2} \left( b c' - b' c \right) = 0,
\label{app:main_a}
\\
 b_{z \bar z} + b X + i \left( a b' - a' b \right) = 0,
\label{app:main_b}
\\
 c_{z \bar z} + c X + i \left( c a' - c' a \right) = 0 ,
\label{app:main_c}
\end{gather}
where
\begin{gather}
  X = a_{z} a_{\bar z}
  + \frac{1}{2} b_{z} c_{\bar z}
  + \frac{1}{2} b_{\bar z} c_{z},
\\
  f' = \lambda^{2} f_{z} + \lambda^{-2} f_{\bar z}
\end{gather}
and
\begin{equation}
a^{2} + bc = 1.
\end{equation}

Consider now the intermediate matrices $\hat V$, $\hat W$ given by
\begin{gather}
\hat V = \partial F \cdot F^{-1} =
  \left(\begin{array}{cc} \hat V_{11} & \hat V_{12} \\ \hat V_{21} & \hat V_{22} \end{array}\right),
\\
\hat W = \bar\partial F \cdot F^{-1} =
\left(\begin{array}{cc} \hat W_{11} & \hat W_{12} \\ \hat W_{21} & \hat W_{22}
\end{array}\right).
\end{gather}
Matrices $V$ (\ref{app:V-psi}) and $W$ (\ref{app:W-psi}), which
are the matrices $V(\lambda)$, $W(\lambda)$ from
Section~\ref{sec_gauge}, can be written as
\begin{equation}
V =
\left(\begin{array}{cc} V_{11} & V_{12} \\ V_{21} & V_{22} \end{array}\right)
=
  \left(\begin{array}{cc}
    \partial \ln D_{11} + \hat V_{11} &
    \displaystyle \frac{ D_{11}}{D_{22}} \hat V_{12} \vspace{1mm}\\
   \displaystyle \frac{ D_{22}}{D_{11}} \hat V_{21} &
    \partial \ln D_{22} + \hat V_{22} \end{array}\right)
\label{app:V}
\end{equation}
and
\begin{equation}
W =
\left(\begin{array}{cc} W_{11} & W_{12} \\ W_{21} & W_{22} \end{array}\right)
=
\left(\begin{array}{cc}    \bar\partial \ln D_{11} + \hat W_{11} &
    \displaystyle\frac{ D_{11}}{D_{22}} \hat W_{12} \vspace{1mm}\\
\displaystyle\frac{ D_{22}}{D_{11}} \hat W_{21} &
    \bar\partial \ln D_{22} + \hat W_{22} \end{array}\right).
\label{app:W}
\end{equation}
Thus, if one takes the matrix $D$ such that its elements satisfy
\begin{equation}
\partial \ln D_{11} = - \hat V_{11},
\qquad
\partial \ln D_{22} = - \hat V_{22}, \label{app:Dii}
\end{equation}
then matrices (\ref{app:V}), (\ref{app:W}) shall have the
following structure:
\begin{equation}
V = \left(\begin{array}{cc} 0 & * \\ * & * \end{array}\right),
\qquad
W = \left(\begin{array}{cc} * & * \\ * & 0 \end{array}\right).
\end{equation}
As to other diagonal elements, they satisfy the identities
\begin{gather}
  \bar\partial V_{22} =
  \bar\partial \hat V_{22} - \partial \hat W_{22},
\\
  \partial W_{11} =
  \partial \hat W_{11} - \bar\partial \hat V_{11}.
\end{gather}
Using (\ref{app:main_a})--(\ref{app:main_c}) one can obtain that
\begin{equation}
  \bar\partial V_{22} =
  \partial W_{11} =
  \frac{ 1}{4a } \left(
    b_{\bar z} c_{z} - b_{z} c_{\bar z}
  \right).
\label{app:ddiag}
\end{equation}

From the definition of the matrices $V$, $W$, $\hat V$, $\hat W$
and $S$ it follows that
\begin{equation}
  V_{12} V_{21} =
  \hat V_{12} \hat V_{21} =
  - \frac{1}{4 } \left( a_{z}^{2} + b_{z} c_{z} \right)
\end{equation}
and
\begin{equation}
  W_{12} W_{21} =
  \hat W_{12} \hat W_{21} =
  - \frac{ 1}{4 } \left( a_{\bar z}^{2} + b_{\bar z} c_{\bar z} \right).
\end{equation}
Using again (\ref{app:main_a})--(\ref{app:main_c}) one can get
that
\begin{gather}
  \bar\partial V_{12} V_{21} =
  - i \lambda^{-2} \bar\partial V_{22},
\\
  \partial W_{12} W_{21} = -i \lambda^{2} \partial W_{11},
\end{gather}
i.e., comparing this result with (\ref{app:ddiag}), one can
conclude that
\begin{gather}
V_{22} = - i \lambda^{2} V_{12} V_{21} + v(z),
\\
W_{11} = i \lambda^{-2} W_{12} W_{21} + w(\bar z).
\end{gather}
Thus, setting $v(z)=-i\lambda^{-2}$, $w(\bar z)=i\lambda^{2}$ and
defying the quantities $q_{0}$, $r_{0}$, $q_{1}$, $r_{1}$ by
\begin{equation}
  q_{0} = - i \lambda^{-1} W_{21},
  \qquad
  r_{0} = i \lambda V_{12},
\qquad
  q_{1} = i \lambda V_{21},
  \qquad
  r_{1} = - i \lambda^{-1} W_{12}
\end{equation}
one can rewrite matrices (\ref{app:V}), (\ref{app:W}) as
\begin{equation}
V =
  \left(\begin{array}{cc} 0 & - i \lambda^{-1} r_{0} \vspace{1mm}\\
   - i \lambda^{-1} q_{1} & - i \lambda^{-2} + i r_{0}q_{1} \end{array}\right),
\qquad
\widetilde V =
\left(\begin{array}{cc} i \lambda^{2} - i q_{0}r_{1} & i \lambda r_{1} \vspace{1mm}\\
    i \lambda q_{0} & 0 \end{array}\right),
\end{equation}
i.e.\ to present them in the form (\ref{V-alh}), (\ref{W-alh}).

To summarize, we have derived, starting from a solution of the
field equation (\ref{main_eq}), the matrix $\Psi$, defined by
(\ref{app:psi-df}), (\ref{app:F}) and (\ref{app:Dii}), which can
be used to perform the gauge transform from the Landau--Lifshitz
linear problems to the ones of the ALH.

\label{pritula-lastpage}
\end{document}